\documentclass[prb,twocolumn,showpacs,groupedaddress]{revtex4}
\usepackage{amsmath}
\usepackage{graphicx}
\usepackage{amsfonts}
\usepackage{amssymb}
\begin{document}
\title{Coherent-incoherent transition in a Cooper-pair-box coupled to a quantum oscillator: an equilibrium approach}
\author{Ying-Hua Huang}
\author{Hang Wong}
\author{Zhi-De Chen}\email[Author to whom correspondence should be addressed: ]{tzhidech@jnu.edu.cn}
\affiliation{Department of Physics, Jinan University, Guangzhou
510632, China}

\begin{abstract}
Temperature effect on quantum tunneling in a Cooper-pair-box coupled
to a quantum oscillator is studied by both numerical and analytical
calculations. It is found that, in strong coupling regions, coherent
tunneling of a Cooper-pair-box can be destroyed by its coupling to a
quantum oscillator and the tunneling becomes thermally activated as
the temperature rises, leading to failure of the Cooper-pair-box.
The transition temperature between the coherent tunneling and
thermo-activated hopping is determined and physics analysis based on
small polaron theory is also provided.

\end{abstract}
 \maketitle

\section{Introduction}
Since the experimental realization of the single-electron transistor
(SET) in 1987,\cite{al,fd} the SET has become an important element
in both extremely precise measurement\cite{bw,zb,kc,la} and quantum
information processing.\cite{aa,mss,dv} The development of the
so-called radio-frequency SET electrometer is considered as the
electrostatic ``dual" of the well known superconducting quantum
interference devices (SQUIDs) and has shown its power of fast and
ultrasensitive measurement.\cite{sch,sch1} On the other hand, the
direct observation of the coherent oscillation in superconducting
SET (SSET, or the so-called Cooper-pair-box (CPB)) shows the SSET
can be served as a physical realization of a coherent two-state
system,\cite{na,na1} and since then the SSET or the SSET-based
circuit has been considered as a candidate of a qubit in a solid
state device and a read-out device of a qubit.\cite{aa,mss,dv}

The high frequency (up to 1 GHz) and high quality factor ($Q\sim
10^3$) nano-mechanical resonator (NR) was firstly fabricated from
bulk silicon crystal in 1996,\cite{cr} and therefore it brought a
lot of interest in demonstrating the quantum nature of this small
mechanical device.\cite{is,ir,lt,ts,ar} A nano-mechanical resonator
capacitively coupled to a SSET forms the so-called
nano-electromechanical system (NEMS), a system shows promises for
fast and ultrasensitive force microscopy.\cite{la,pb} Especially,
displacement-detection approaching the quantum limit by this system
has been demonstrated.\cite{la} From a view point of theoretical
study, NEMS can be interesting in their own right. In fact, NEMS
provides a physical realization of a simple quantum system: a
two-level system (TLS) coupled to a single phonon mode, which has
basic relation to some important models in solid state physics and
quantum optics. First of all, it represents an oversimplified
spin-boson model,\cite{leg,weiss} i.e., the spin-boson model in a
single-phonon mode case, also it can be considered as the Einstein
model in a two-site problem in the exciton-phonon (or
polaron-phonon) system.\cite{ss} By taking the analog of the
resonator as the single mode cavity, it is the Jaynes-Cummings model
in quantum optics.\cite{lt,jm,jc,sun}

To be a physical realization of a TLS, the SSET needs to work in
severe constrains. It is well known that both thermal and quantum
fluctuations from the environment can destroy the Coulomb blockade
of tunneling and lead to the failure of the SET.\cite{al1,hd}
Additional conditions are needed for a SSET,\cite{al1} and one
important factor is to maintain the coherent tunneling. The study on
polaron-phonon system shown that the coherent motion of the electron
can be destroyed by its coupling to the phonon and a transition from
band motion to hopping motion happens as the temperature
rises.\cite{mah} By analogy, it is important to know if the coherent
tunneling of the SSET in NEMS can be destroyed by its coupling to a
NR, leading to failure of the SSET. We shall state such a
phonon-induced transition as coherent-incoherent transition and this
is the main interest of the present paper.

The coherent-incoherent transition in NEMS will be studied by both
numerical and analytical analyses. It is shown that the
coherent-incoherent transition exists for some frequency ranges with
strong coupling parameters. The rest of the present paper is
organized as follows. In Sec.\ II, we explain the model and the way
to trace out the environment, the results from adiabatic
approximation is also discussed. The coherent-incoherent transition
is studied in Sec.\ III and conclusion and discussion are presented
in Sec.\ IV.

\section{The model and explanation of the approximation}
The Hamiltonian of the NEMS, i.e., a TLS $+$ a NR, is given by
(setting $\hbar=1$)\cite{mss,ar}
\begin{equation}
\hat{H}_0=\frac{1}{2}\epsilon_0\hat{\sigma}_z-\frac{1}{2}\Delta_0\hat{\sigma}_x+\Omega\hat{a}^{\dagger}\hat{a}+\lambda\hat{\sigma}_z(\hat{a}^{\dagger}+\hat{a}),
\end{equation}
where $\epsilon_0=E_C(1-2n_g)$ and $E_C$ is the charging energy of a
Cooper pair, $n_g=C_gV_g/(2e)$ with $C_g$ and $V_g$ are the gate
capacity and gate voltage, respectively. $\Delta_0=E_J=I_c/(2e)$ and
$I_c$ is the critical current of the Josephson junction.
$\hat{\sigma}_i$ ($i=x,y,z$) are the Pauli matrices and
$\hat{a}^{\dagger}$  $(\hat{a})$ is the creation (annihilation)
operator of the phonon mode with energy $\Omega$, while $\lambda$ is
the coupling parameter. The above Hamiltonian indicates that our
main interest is the case when the SSET is closed to the degeneracy
point.\cite{mss,ar} It should be noted that the system described by
the above Hamiltonian is not a thermodynamical system, a true
thermodynamical system should include the environment, which is
modeled as a collection of harmonic oscillators.\cite{leg,weiss} The
whole Hamiltonian can be written as
\begin{equation}
\hat{H}=\hat{H}_0+\hat{H}_e,\quad\hat{H}_e=\sum_k\omega_k\hat{b}_k^{\dagger}\hat{b}_k+\hat{H}_{int}, 
\end{equation}
where $\hat{H}_{int}$ represents the coupling between the
environment and NEMS. In the present paper, we shall assume that the
effect of the environment is just to keep the NEMS in equilibrium.
Here we apply the concept of reduced density matrix to explain the
approximation. The density matrix of the whole system is
\begin{equation}
\rho=\rho_0\otimes\rho_e,
\end{equation}
where $\rho_0$ and $\rho_e$ are the density matrix of the NEMS and
environment, respectively. The expectation value of any operator
$\hat{Q}$ that acts only on the variables of the NEMS can be found
by\cite{bl}
\begin{equation}
\langle\hat{Q}\rangle={\rm
Tr}\rho_r(\phi)\langle\phi|\hat{Q}|\phi\rangle,\quad \rho_r={\rm
Tr}_{bath}\rho,
\end{equation}
where $\rho_r$ is the reduced density matrix and ``${\rm
Tr}_{bath}$" means the trace operation over the environment. Our
approximation means that the above expression can be rewritten as
\begin{equation}
\langle\hat{Q}\rangle\simeq \frac{1}{Z}\sum_ie^{-\beta E_i}\langle\phi_i|\hat{Q}|\phi_i\rangle,~~~~~~~~
\hat{H}_0|\phi_i\rangle=E_i|\phi_i\rangle,
\end{equation}
with $Z=\sum_ie^{-\beta E_i}$ is the partition function of the
NEMS and $\beta=1/T$ (setting $k_B=1$). In previous
treatments,\cite{ir,lt} the effect of the environment is just to
keep the NR in equilibrium. The present treatment, taking a small
step further, suggests that the NEMS is in equilibrium with the
environmental bath.

By taking this approximation, thermodynamical properties of the NEMS
can be known, provided that the eigenvalue problem of $\hat{H}_0$ in
Eq.\ (1) can be solved. The eigenvalues of $\hat{H}_0$ can be obtain
by the standard numerical diagonalization technique. In a practical
treatment, the number of eigenvalues treated must be finite, we
therefore truncate the eigenspectrum by just taking the first
smallest $N$ eigenvalues, i.e.,
\begin{equation}
\langle\hat{Q}\rangle\simeq \frac{1}{Z}\sum_ie^{-\beta E_i}\langle\phi_i|\hat{Q}|\phi_i\rangle
\simeq \frac{1}{Z}\sum_{i=1}^Ne^{-\beta E_i}\langle\phi_i|\hat{Q}|\phi_i\rangle,
\end{equation}
with $Z\simeq \sum_{i=1}^Ne^{-\beta E_i}$, while $N$ is determined by the following condition
\begin{equation}
\beta(E_N-E_1)\ge L,
\end{equation}
where $L$ is a fixed number to control the calculation error and in
the present paper we have $L\ge 20$. Accordingly, one can find the
interested physical quantities from the numerical results, for
example, the free energy of the NEMS is
\begin{equation}
F(T)\simeq  -T\ln\left(\sum_{i=1}^Ne^{-\beta E_i}\right),
\end{equation}
from which all the thermodynamical quantities of the NEMS can be
found. In our numerical calculation, we have randomly checked the
dependence of the results on the value of $L$. We found no
observable dependence on $L$ for $L\ge 20$ by extending $L$ to 50 or
even 100. It should be noted that, for calculating the temperature
dependence of tunneling splitting, Eq.\ (5) should be modified as
\begin{equation}
\Delta(T)\simeq \frac{1}{Z}\sum_{i=1}^Ne^{-\beta E_i}|\langle\phi_i|\hat{\sigma}_x|\phi_i\rangle|,
\end{equation}
this is because the sign of $\langle\phi_i|\hat{\sigma}_x|\phi_i\rangle$ is not relevant in the calculation of $\Delta(T)$.

Before going to numerical results, we shall first discuss the
adiabatic approximation in the case of $\epsilon_0=0$ which was
shown to be a good approximation for $\Delta_0/\Omega \le
1$.\cite{ir} In the adiabatic approximation, the eigenvalues of
$\hat{H}_0$ are characterized by pairs of energies that are given
by\cite{ir}
\begin{equation}
E_a^{\pm}(n)=n\Omega\pm\Delta_a(n)-\frac{\lambda^2}{\Omega},\quad\Delta_a(n)=\frac{\Delta_0}{2}\langle
n_+|n_-\rangle,
\end{equation}
where $|n_{\pm}\rangle=\exp\{\mp(\lambda/\Omega)(a^+-a)\}|n\rangle$
are the displaced-oscillator states and $n=0,1,2,\cdots$. Taking
these eigenvalues, one can find the temperature dependence of the
tunneling splitting according to Eq.\ (9) and the result is
\begin{equation}
\Delta_a(T)\simeq \frac{1}{Z_a}\sum_{n=0}^{\infty}\Delta_a(n)
e^{-\beta n\Omega}~2\cosh(\beta\Delta_a(n)),
\end{equation}
where $Z_a=\sum_{n=0}^{\infty}e^{-\beta
n\Omega}~2\cosh(\beta\Delta_a(n))$. In the limit of
$\Delta_a(n)\ll\Omega$, it can be found that
\begin{equation}
\Delta_a(T)\simeq \sum_{n=0}^{\infty}\Delta_a(n) p_{{\rm th}}(n),~~~~~~p_{{\rm th}}(n)=e^{-\beta n\Omega}(1-e^{-\beta \Omega}),
\end{equation}
a result that was used in the previous treatments.\cite{ir,lt}
\section{Coherent-incoherent transition}
To perform numerical diagonalization, one needs to represent the
Hamiltonian $\hat{H}_0$ with a suitable basis. In the treatment of
Ref.\ \onlinecite{ir}, the basis of $|\uparrow,\downarrow\rangle
\otimes |n_{\pm}\rangle$ is applied, where $|\uparrow\rangle$ and
$|\downarrow\rangle$ are the eigenstates of $\sigma_z$. However,
with this basis, the elements of the Hamiltonian matrix are flooded
with the overlap terms between different states $|n_\pm\rangle$ as
shown in the Eq.\ (8) of Ref.\ \onlinecite{ir}. Such overlap terms
are rather complicated to compute and therefore the generation of
the Hamiltonian matrix will be slow. In our treatment, for
convenient, we use a very simple basis that consist of the
eigenstates of $\sigma_z$ and $\hat{a}^\dagger\hat{a}$ to represent
the Hamiltonian, i.e., a basis of $|\uparrow,\downarrow\rangle
\otimes |n\rangle$. Hence, we represent the TLS terms with
$|\uparrow,\downarrow\rangle$ and the NR terms with $|n\rangle$ and
do a tensor product, the matrix of $\hat{H}_0$ is generated. Of
course, both schemes give the same results. The diagonalization can
give the eigenvalues of $\hat{H}_0$, i.e., $\{E_i\}$, and the
corresponding eigenvectors $\{\phi_i\}$, from which the tunneling
splitting $\langle\phi_i|\hat{\sigma}_x|\phi_i\rangle$ can be found.
Noted that, to reach the accuracy condition (7), the matrix size is
$\Omega$-dependent, the lower the frequency, the larger the matrix
size. In the following, we shall firstly concentrate our interest on
the unbiased case of $\epsilon_0=0$ and the biased case will be
discussed in the end of this section. The eigenvalues in the case of
$\epsilon_0=0$ we found are in good agreement with the results in
Ref.\ \onlinecite{ir}, i.e., the results of the adiabatic
approximation given in Eq.\ (10) work well for $\Omega/\Delta_0\ge
1$. The corresponding tunneling splitting of the eigenstates for
$\lambda/\Omega=0.5$ and some typical values of $\Delta_0/\Omega$
are shown in Fig.\ 1, where the results of the adiabatic
approximation are also given for comparison.

\begin{figure*}
\centering
\begin{minipage}[c]{0.7\textwidth}
\centering
\includegraphics[width=\textwidth]{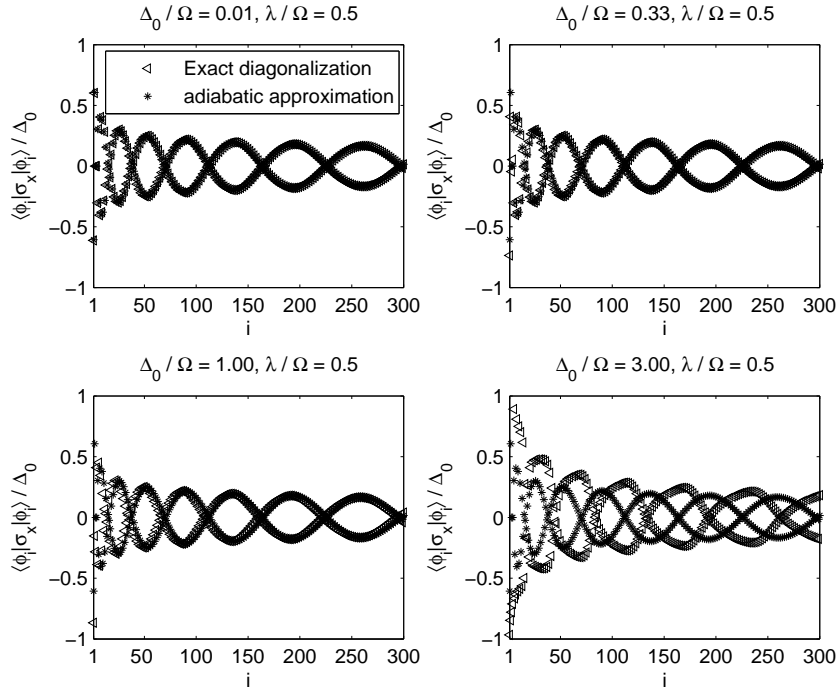}
\end{minipage}
\begin{minipage}[c]{0.25\textwidth}
\centering \caption{Tunneling splitting for eigenstates of
$\hat{H}_0$ in the case of $\lambda/\Omega=0.5$ and $\epsilon_0=0$
for some typical values of $\Delta_0/\Omega$ by numerical
calculation (shown as triangles). The results of the adiabatic
approximation (shown as ``$\ast$") are also shown for comparison.
The sketches of both curves show good agreement for
$\Delta_0/\Omega\le 1$ and the higher the excited states the better
the agreement. However, obvious discrepancy appears for both the
ground state and the first excited state even when $\Delta_0/\Omega=
1/3$. The results of the adiabatic approximation show large
discrepancy with the numerical results in low frequency region with
$\Delta_0/\Omega> 1$.}
\end{minipage}%
\end{figure*}

\begin{figure}[h]
\centering
\includegraphics[width=0.5\textwidth]{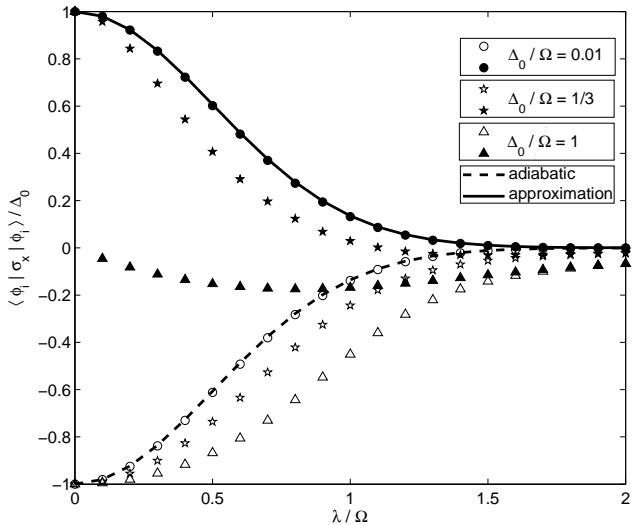}
\caption{Comparison of the tunneling splitting of both the ground
state and the first excited state by numerical calculation with
adiabatic approximation in the case of $\epsilon_0=0$ for some
typical values of $\Delta_0/\Omega$. It should be noted that the
result by adiabatic approximation is
$\Delta_0/\Omega$-independent, which is shown as solid and dash
lines (dash line is the ground state). Numerical results are shown
in different patterns for different values of $\Delta_0/\Omega$
(the hollow patterns are for the ground state).}
\end{figure}

It can be found that the sketches of both curves show good agreement
for $\Delta_0/\Omega\le 1$ and the higher the excited states the
better the agreement. However, a closed look at the curves shows
that obvious discrepancy appears for both the ground state and the
first excited state even when $\Delta_0/\Omega= 1/3$. Figure 2 shows
the details of the discrepancy. We found that the discrepancy begins
to appear even when $\Delta_0/\Omega=0.1$ and becomes obvious when
$\Delta_0/\Omega = 1/3$. The present results show that, if one
compares the eigenvalues of the adiabatic approximation with the
numerical diagonalization, the adiabatic approximation seems to work
pretty well for $\Delta_0/\Omega\le 1$; however, if one measures the
tunneling splitting, the adiabatic approximation works well only for
$\Delta_0/\Omega\le 0.1$.

After finding out the eigenvalues and the corresponding tunneling
splitting, one can calculate temperature dependence of the tunneling
splitting according to Eq.\ (8). For all the frequencies we studied
(down to $\Delta_0/\Omega=50$), it is found that the tunneling
splitting decreases with temperature at the beginning in low
temperature regions. However, when the coupling strength
$\lambda/\Omega$ increases to some critical value, $\Delta(T)$ has a
``upturn" at some temperature $T_t$, which is
$\lambda/\Omega$-dependent. Typical results are shown in Fig.\ 3.

\begin{figure*}
\centering
\begin{minipage}[c]{0.74\textwidth}
\centering
\includegraphics[width=\textwidth]{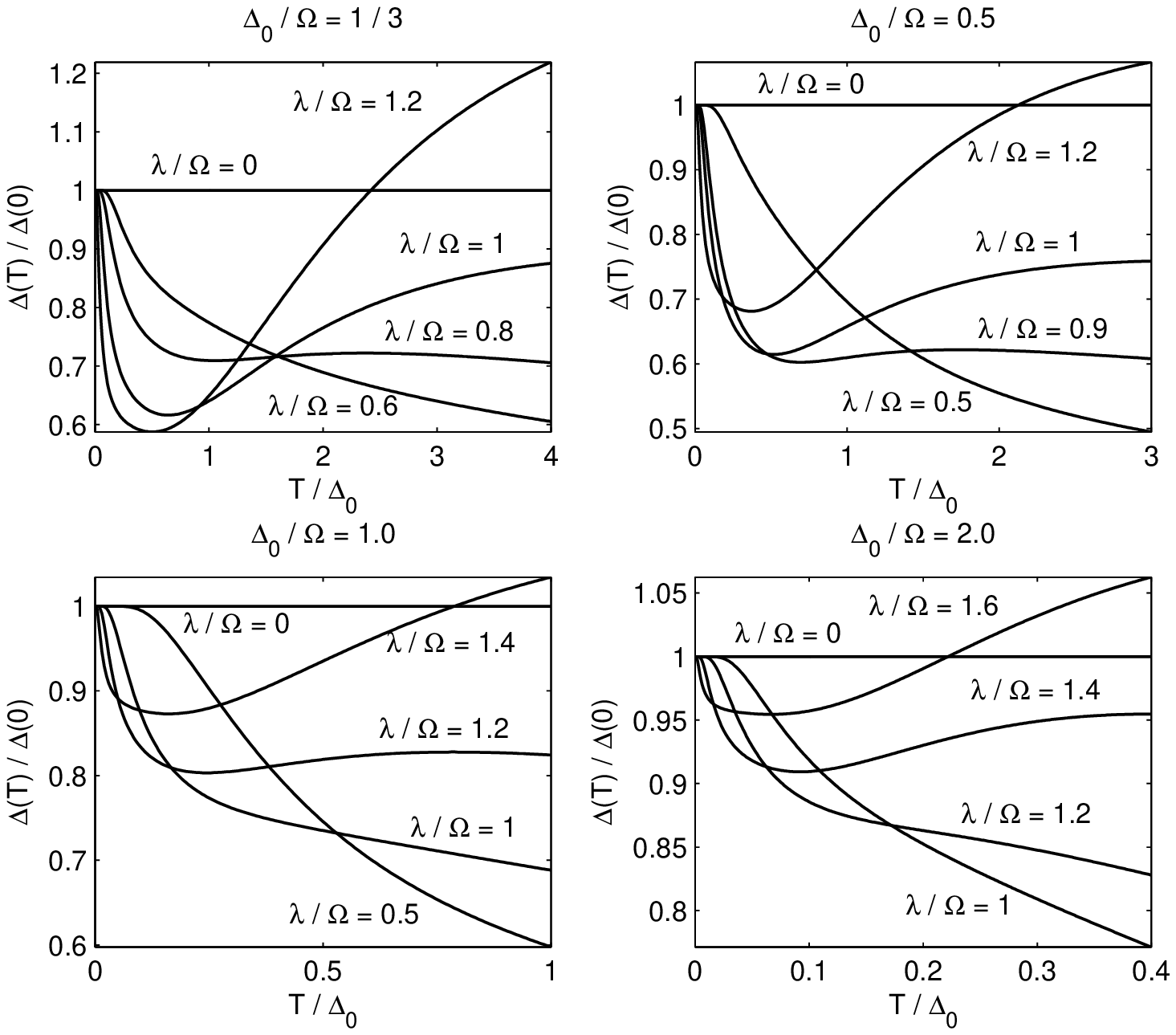}
\end{minipage}
\begin{minipage}[c]{0.25\textwidth}
\centering \caption{Temperature dependence on the tunneling
splitting in the case of $\epsilon_0=0$ for some typical values of
$\Delta_0/\Omega$. At the beginning, $\Delta(T)$ decreases with
temperature, but as the coupling parameter  $\lambda/\Omega$
increases, $\Delta(T)$ shows a ``up-turn" at some temperature $T_t$,
which is $\lambda/\Omega$-dependent.}
\end{minipage}%
\end{figure*}

The curves shown in Fig.\ 3 indicate that, for a given value of
$\Delta_0/\Omega$ and in the low temperature region, the main effect
of the NR on quantum tunneling is to decreases the tunneling
splitting in the weak coupling regions. The situation is more or
less the same as that in the polaron-phonon (or exciton-phonon)
system, say, coupling to phonons makes the electron become a
``dressed" one and lowers the hopping rate of the electron. However,
as the coupling strength increases, the effect of the NR on quantum
tunneling changes as temperature increases to $T_t$, e.g., the
tunneling splitting is enhanced by a NR as temperature increases
further. The result is reminiscent of the transition from Bloch-type
band motion to phonon-activated hopping motion in the polaron-phonon
(or exciton-phonon) system.\cite{mah} Moreover, $\Delta_a(T)$
obtained from Eq.\ (11) does not show such ``upturn" for the
frequency regions we studied. This implies that the ``up-turn" is a
non-adiabatic effect.

As we have mentioned in Sec. I, the present model can be considered
as an analog to the polaron-phonon system of
Einstein model.\cite{mah} We therefore employ the small polaron theory to analyze the temperature dependence of the tunneling splitting. 
At the beginning of the low temperature regions, the expectation
value of phonon number is small, the main contribution to the
tunneling splitting is the so-called diagonal transitions. The
diagonal transition rate can be found by following the way given in
Ref.\ \onlinecite{mah}. Firstly, a canonical transformation is
applied to the Hamiltonian in Eq.(2), i.e.,
\begin{equation}
\hat{H}'=e^S\hat{H}e^{-S}=\hat{{\cal H}}_0+\hat{V}+\sum_k\omega_k\hat{b}_k^{\dagger}\hat{b}_k+\hat{H}_{int}',
\end{equation}
where $S=(\lambda/\Omega)\hat{\sigma}_z(\hat{a}^{\dagger}-\hat{a})$,
\begin{equation}
\hat{{\cal H}}_0=\frac{\epsilon_0}{2}\hat{\sigma}_z+\Omega\hat{a}^{\dagger}\hat{a}-\frac{\lambda^2}{\Omega},
\end{equation}
\begin{equation}
\hat{V}=\frac{\Delta_0}{4}(\hat{\sigma}_+e^{(\lambda/\Omega)(\hat{a}-\hat{a}^{\dagger})}+h.c.),
\end{equation}
$\hat{H}_{int}'=e^S\hat{H}_{int}e^{-S}$ and $\hat{\sigma}_{\pm}=\hat{\sigma}_x\pm i\hat{\sigma}_y$. The diagonal transition rate is given by
\begin{equation}
w_d(T)={\rm Tr}_{\hat{{\cal H}}_0}\langle i|\hat{V}|i\rangle=\frac{\Delta_0}{2}e^{-S_T},
\end{equation}
where $S_T=(2\lambda/\Omega)^2(n+1/2)$ and
$n=(e^{\beta\Omega}-1)^{-1}$ is the expectation value of phonon
number. Here, we have employed the approximation elucidated in Sec.
II, i.e., the effect of
$\sum_k\omega_k\hat{b}_k^{\dagger}\hat{b}_k+\hat{H}_{int}'$ is just
to keep the NEMS in equilibrium. $w_d$ and hence the tunneling
splitting will decrease with increasing temperature. On the other
hand, the main contribution to tunneling splitting, at higher
temperature, is the so-called non-diagonal transitions. One can
calculate the correlation function in the way given in Ref.\
\onlinecite{mah}, e.g.,
\begin{equation}
W(t)={\rm Tr}_{\hat{{\cal H}}_0}[\langle
i|\hat{V}(t)\hat{V}(0)|i\rangle-|\langle i|\hat{V}|i\rangle|^2],
\end{equation}
with $\hat{V}(t)=e^{i\hat{{\cal H}}_0t}\hat{V}e^{-i\hat{{\cal H}}_0t}$ and the result is
\begin{equation}
W(t)=(\Delta_0/2)^2e^{-2S_T}[e^{\varphi(t)}-I_0(\epsilon)],
\end{equation}
where $I_0(x)$ is a Bessel function and
\begin{equation}
\varphi(t)=\epsilon \cos[\Omega(t+i\beta\hbar/2)],\quad
\epsilon=2(2\lambda/\Omega)^2[n(n+1)]^{1/2}.
\end{equation}
It is interesting to note that, in the present case, the factor
$I_0(\epsilon)$ still survives after taking thermodynamical limit, a
situation that is different from the small polaron theory of the
Einstein model.\cite{mah} This is because only one phonon mode
coupled to the TLS in the present case and all the other modes just
serve as the bath. This result can also help to get rid of the delta
function problem shown in Ref.\ \onlinecite{mah}. The non-diagonal
transition rate can be found by
using the saddle-point integration
\begin{equation}
w_n(T)=\int_{-\infty}^{\infty}W(t)dt\simeq (\Delta_0/2)^2(\pi/\gamma)^{1/2}e^{-2S_T+\epsilon},
\end{equation}
where $\gamma=\lambda^2[n(n+1)]^{1/2}$. It can be easily checked that $w_n$ shows opposite temperature dependence to $w_d$ and the transition
temperature $T_t$ is determined by $w_d=w_n$, which leads to
\begin{equation}
\frac{\Delta_0}{2}[\pi/\gamma(T_t)]^{1/2}e^{\epsilon(T_t)-S_T(T_t)}=1.
\end{equation}
The above equation can be solved numerically and the transition
temperatures obtained are shown in Fig.\ 4, where the transition
temperatures determined from $\Delta(T)$ by numerical calculation
(i.e., curves shown in Fig.\ 3) are also presented for comparison.

\begin{figure*}
\centering
\begin{minipage}[c]{0.74\textwidth}
\centering
\includegraphics[width=\textwidth]{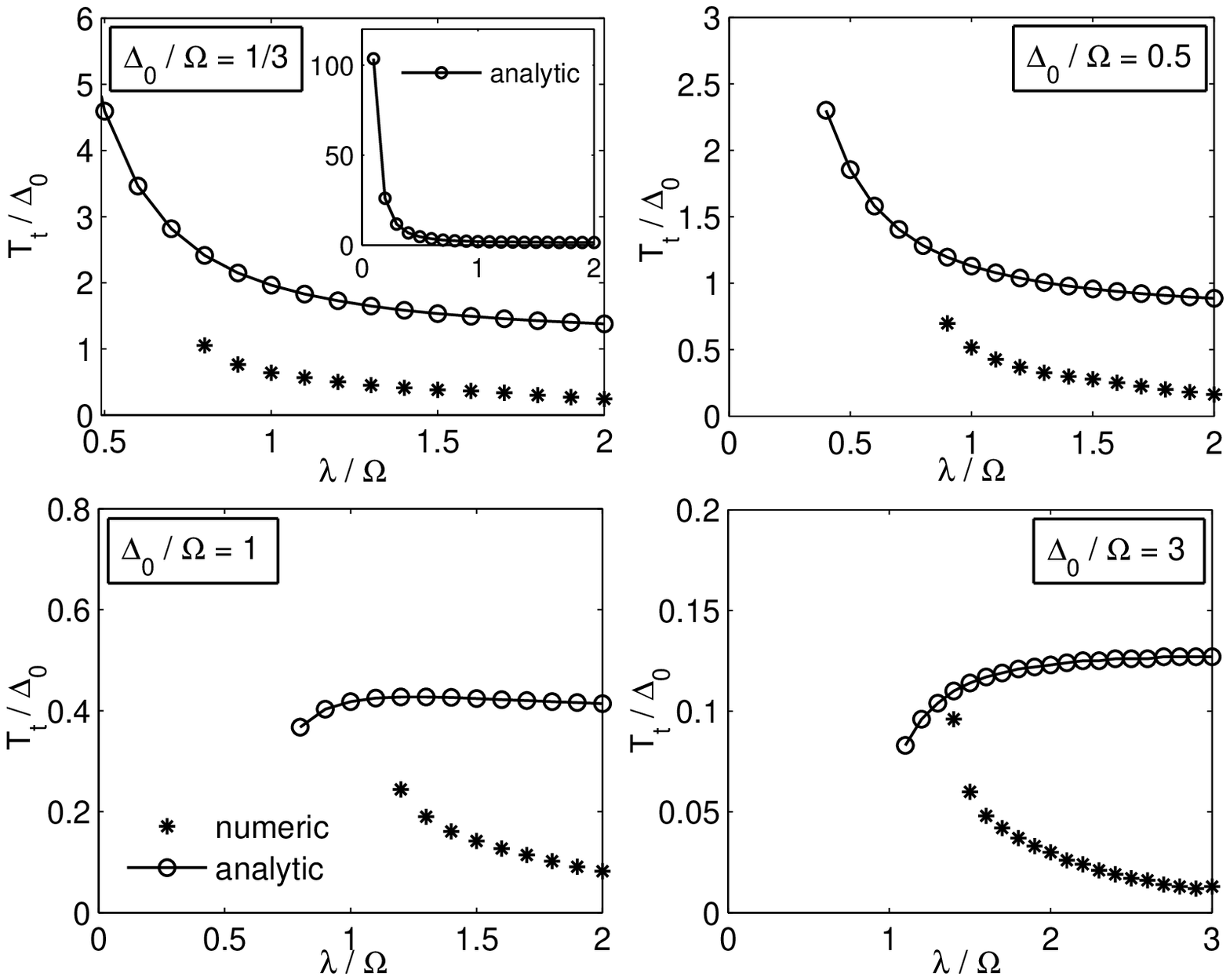}
\end{minipage}
\begin{minipage}[c]{0.25\textwidth}
\centering \caption{Transition temperatures obtained by analytical
(from Eq.(21)) and numerical calculations in the case of
$\epsilon_0=0$ for some typical values of $\Delta_0/\Omega$. The
analytical results are always higher than the numerical ones. For
$\Delta_0/\Omega<1$, both transition temperatures show similar
$\lambda/\Omega$ dependence, i.e., $T_t$ decreases with increasing
$\lambda/\Omega$. However, opposite $\lambda/\Omega$ dependence is
found for $\lambda/\Omega\ge 1$, showing that the analytical results
are invalid in the low frequency regions. The inset of the lefttop
figure shows the whole transition temperature curve obtained by the
analytical calculation.}
\end{minipage}%
\end{figure*}

It can be seen from Fig.\ 4 that the transition temperatures from
numerical calculations are always lower than the analytical ones. In
high frequency region with $\Delta_0/\Omega<1$, both obtained
transition temperatures show similar $\lambda/\Omega$ dependence,
i.e., $T_t$ increases as $\lambda/\Omega$ decreases. However, the
transition temperature obtained from analytical calculation shows
opposite coupling parameter dependence in low frequency region with
$\Delta_0/\Omega<1$, a result that is in confliction with an
intuitive picture. We notice that, as shown in the inset of Fig.\ 4,
analytical calculation predicts a transition for coupling strength
as low as $\lambda/\Omega\sim 0.2$ for $\Delta_0/\Omega\le 1/3$;
however, numerical result shows that there is no transitions for
coupling strength $\lambda/\Omega$ lower than 0.7. Figure 5 draws
the the transition boundary as a function of $\lambda/\Omega$ and
$\Delta_0/\Omega$ obtained by the numerical calculation.

\begin{figure}
\centering
\includegraphics[width=0.5\textwidth]{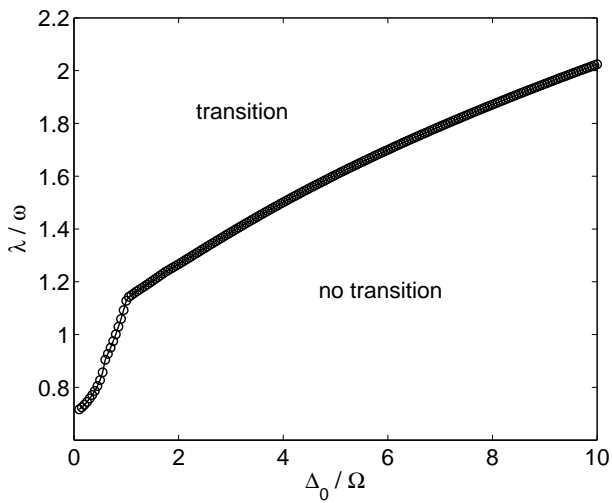}
\caption{Transition boundary obtained by the numerical calculation
in the case of $\epsilon_0=0$. The larger the $\Delta_0/\Omega$
(i.e., the lower frequency for a given $\Delta_0$), the larger the
coupling strength $\lambda/\Omega$ is needed for the transition to
happens, indicating the SSET is more stable when coupling to a lower
frequency NR.}
\end{figure}

We believe that the discrepancy mainly comes from the approximation
made in the analytical calculation of the diagonal transition rate.
It can be shown that
\begin{equation}
w_d(T)=\sum_{n=0}^{\infty}\Delta_a(n) p_{{\rm th}}(n)\simeq \Delta_a(T),
\end{equation}
which indicates that effectively the diagonal transition rate is
found by adiabatic approximation, an approximation is good for
finding the tunneling splitting only when $\Delta_0/\Omega\le 0.1$.
Figure 6 shows the details of how the discrepancy comes from the
adiabatic approximation.

\begin{figure}
\centering
\includegraphics[width=0.5\textwidth]{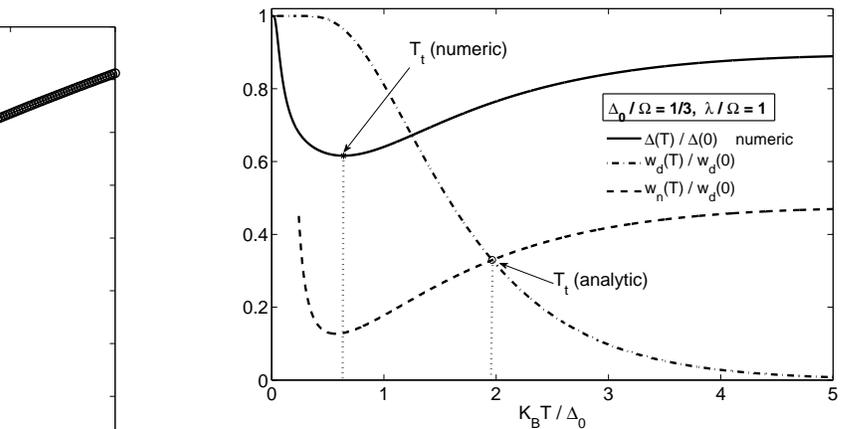}
\caption{Illustration of how the discrepancy comes from $w_d(T)$
obtained by the adiabatic approximation in the case of
$\epsilon_0=0$, $\Delta_0/\Omega=1/3$, and $\lambda/\Omega=1$. As
the temperature increases, the descending rate of $w_d(T)/w_d(0)$ is
much lower than $\Delta(T)/\Delta(0)$ from the numerical
calculation, leading to a higher transition temperature. The
``up-turn" of $w_n(T)$ in the low temperature regions is due to the
factor $\gamma^{-1/2}$ which diverges as $T\rightarrow 0$. }
\end{figure}

It turns out that, as the temperature increases, the descending rate
of $w_d(T)/w_d(0)$ is much lower than $\Delta(T)/\Delta(0)$ from the
numerical calculation. Such a result has two consequences. The first
one is the transition temperature by analytical calculation is
higher than the numerical result since a slower descending
$w_d(T)/w_d(0)$ will lead to a higher crosspoint with a given
$w_n(T)$; Still another is the extremely high transition temperature
in the weak coupling regions shown in the inset of Fig.\ 4. It is
found that, in the weak coupling regions, the slow descending
$w_d(T)/w_d(0)$ can still survive to some high temperature, at where
$\Delta(T)$ from the numerical calculation has died out, bringing
about an artifact high transition temperature. On the other hand, As
one can see from Fig.\ 1 and Fig.\ 2, tunneling splitting from the
adiabatic approximation shows large discrepancy with the numerical
result when $\Delta_0/\Omega\ge 1$. This implies that $w_d(T)$ is a
bad approximation to calculate the diagonal transition rate in low
frequency regions with $\Delta_0/\Omega\ge 1$. Numerical analysis
indicates that $w_d(T)$ shows different
$(\lambda/\Omega)$-dependence from $\Delta(T)$, leading to different
$\lambda/\Omega$-dependence of $T_t$. In other words, the break-down
of the adiabatic approximation in the low frequency region with
$\Delta_0/\Omega\ge 1$ leads to an abnormal
$\lambda/\Omega$-dependence of the transition temperature in the low
frequency regions.

Now we turn to see the effect of bias on the coherent-incoherent
transition. In the case of $\epsilon_0\not=0$, intuitively, the
hopping between the TLS needs the assistance of the phonon unless
the bias is very small comparing with $\Delta_0$, i.e., the bias
can be overcome by the quantum fluctuation. Accordingly, the
diagonal transition is still possible and hence the
coherent-incoherent transition is expected to survive only when
$\epsilon_0/\Delta_0\ll 1$. Nevertheless, the appearance of the
non-zero bias will weaken the diagonal transition rate. As
$\epsilon_0$ increases to some value, at where tunneling becomes
impossible without the assistance of phonon, then the diagonal
transition makes no contribution to tunneling and
coherent-incoherent transition disappears. Numerical calculation
in the case of $\epsilon_0\not=0$ is the same as $\epsilon_0=0$
and some typical results obtained by the numerical calculation are
shown in Fig.\ 7. As $\epsilon_0/\Delta_0$ increases, the
transition temperature decreases and the transition disappears
when $\epsilon_0/\Delta_0$ reaches some critical value which is a
function of $\lambda/\Omega$ and $\epsilon_0/\Omega$. Obviously,
the numerical result is in accord with the analysis presented
above.

\begin{figure*}
\centering
\begin{minipage}[c]{0.74\textwidth}
\centering
\includegraphics[width=\textwidth]{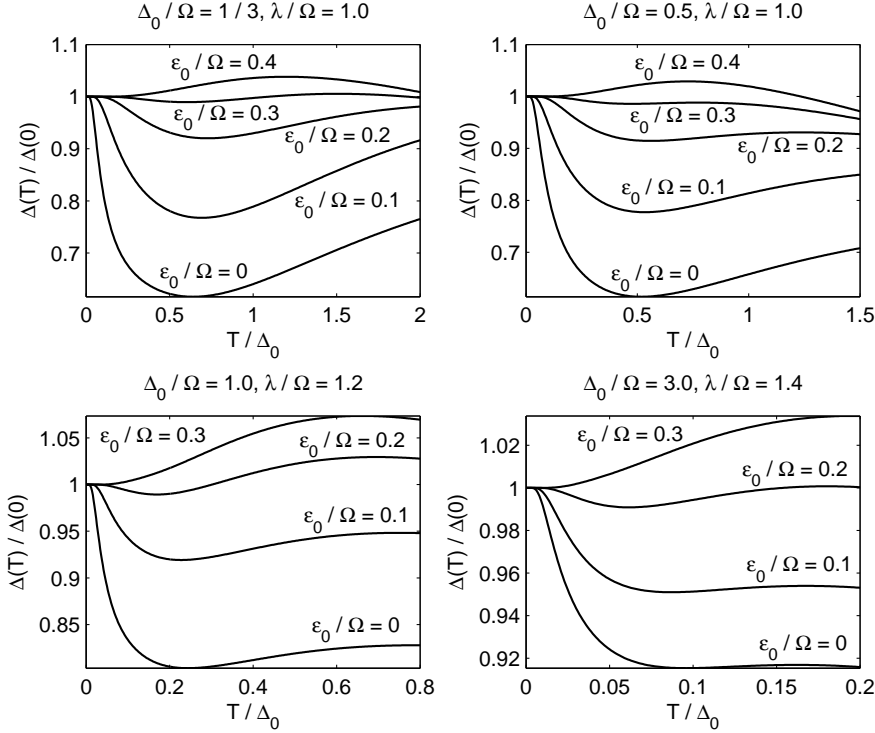}
\end{minipage}
\begin{minipage}[c]{0.25\textwidth}
\centering \caption{The effect of bias on the coherent-incoherent
transition for some typical values of $\lambda/\Omega$ and
$\Delta_0/\Omega$. The transition temperature decreases with
increasing $\epsilon_0/\Omega$ and the transition disappears when
$\epsilon_0/\Omega$ reaches some critical value which depends on
both $\lambda/\Omega$ and $\Delta_0/\Omega$.}
\end{minipage}%
\end{figure*}

\section{Conclusion and discussion}
In conclusion, we have presented study on the effect of a
nanomechanical resonator on quantum tunneling in a Cooper-pair-box
at $T\not=0$. We found that the coherent tunneling of a
Cooper-pair-box can be destroyed by the coupling to a NR at a
temperature much lower than the coupling energy $E_J$ of the
Josephson junction. The present analysis shows that, for a NEMS to
work well, one additional condition, e.g., $T\ll T_t$, is needed.
The coherent-incoherent transition boundary as a function of
$\Delta_0/\Omega$ and $\lambda/\Omega$ is calculated. It turns out
that the transition happens only for $\lambda/\Omega>0.7$, and the
lower the $\Delta_0/\Omega$, the larger the $\lambda/\Omega$ is
needed for the transition. We also found that the transition
temperature is a monotonic descending function of
$\lambda/\Omega$. The present result shows that experimental
observation on the coherent-incoherent transition is still
impossible since the corresponding coupling parameter cannot be
achieved in the present stage. Taking some typical experimental
parameters:\cite{is} $\Delta_0=4\mu$eV and $\hbar\Omega=1.2\mu$eV,
the coupling parameter needed for the transition is
$\lambda/\Omega\simeq 1.4$, which is larger than the strong
coupling limit possible to achieve in experiment
$(\lambda/\hbar\omega_0\sim 1)$.\cite{ir} To see the transition in
the coupling parameters can be achieved in experiment, the result
shown in Fig.\ 5 tells $\Delta_0/\Omega$ should lower than 1,
i.e., the frequency of the NR should be as high as 1GHz, which is
almost the limitation in the present stage. Nevertheless, it is
believed that, in a real system, the transition can be seen in a
lower coupling parameter regions since the coupling of the NR to
the environment can help the transition to happen. It is also
found that other thermodynamical quantities, like specific heat,
vary smoothly over the transition point, showing that the
transition is not a thermodynamical transition.

Coherent-incoherent transition is an important issue in the
spin-boson model and analysis on the transition at $T\not=0$ was
also provided in Ref.\ \onlinecite{leg}. However, the starting
point in Ref.\ \onlinecite{leg} is different from the present
analysis. In the present analysis, the key element is the
temperature dependence on the tunneling splitting, while in the
previous analysis, it is the relaxation behavior, i.e., the time
dependence on transition rate $P(t)$.\cite{leg} As a matter of
fact, the present analytic calculation is similar to the
calculation provided in Sec. III D of Ref.\ \onlinecite{leg}, and
accordingly the transition rate we found here is corresponding to
$\Gamma$ (or $1/(2\tau)$), which was predicted to have a monotonic
temperature dependence in their analysis. It seems that this
result is suitable for the case with large bias while the
``up-turn" of $\Delta(T)$ for small (or zero) bias cannot be
explained. Nevertheless, the present result can help to understand
the coherent-incoherent transition at $T\not=0$ in the spin-boson
model with small bias. To the first order approximation (i.e.,
omitting the cooperative effect between different phonon modes),
the combined contribution of all the phonon modes with a
frequency-dependent weight for $0 <\omega<\omega_c$ is
approximately the contribution of the whole bath, accordingly a
coherent-incoherent transition is expected to exist in the
spin-boson model since the coupling of the TLS to all the phonon
modes can lead to the transition when the coupling strength
exceeds some value.

\begin{acknowledgements}
This work was supported by a grant from the Natural Science
Foundation of China  under Grant No. 10575045.
\end{acknowledgements}

\end{document}